\documentclass[11pt,a4paper]{article}

\usepackage{fullpage}
\usepackage{amsmath}
\usepackage{amssymb}
\usepackage{mathrsfs}
\usepackage{graphicx}
\usepackage{psfrag}
\usepackage[all,poly,curve,knot,cmtip]{xy}
\usepackage{citesort}

\newcommand {\be} {\begin{eqnarray*}}
\newcommand {\ee} {\end{eqnarray*}}
\newcommand {\bea} {\begin{eqnarray}}
\newcommand {\eea} {\end{eqnarray}}

\newcommand{\bm}[1]{\boldsymbol{#1}}

\begin{document}

\title{Supersymmetric isolated horizons}
\author{\textbf{Tom$\acute{\mbox{a}}\check{\mbox{s}}$
Liko}\footnote{Electronic mail: tliko@math.mun.ca}\\
\\{\small \it Department of Physics and Physical Oceanography}\\
{\small \it Memorial University of Newfoundland}\\
{\small \it St. John's, Newfoundland, Canada, A1B 3X7}\\
\\\textbf{Ivan Booth}\footnote{Electronic mail: ibooth@math.mun.ca}\\
\\{\small \it Department of Mathematics and Statistics}\\
{\small \it Memorial University of Newfoundland}\\
{\small \it St. John's, Newfoundland, Canada, A1C 5S7}}

\maketitle





\begin{abstract}

We construct a covariant phase space for rotating weakly isolated
horizons in Einstein-Maxwell-Chern-Simons theory in all (odd)
$D\geq5$ dimensions.  In particular, we show that horizons on the
corresponding phase space satisfy the zeroth and first laws of
black-hole mechanics.  We show that the existence of a Killing
spinor on an isolated horizon in four dimensions (when the
Chern-Simons term is dropped) and in five dimensions requires
that the induced (normal) connection on the horizon has to vanish,
and this in turn implies that the surface gravity and rotation
one-form are zero.  This means that the gravitational component
of the horizon angular momentum is zero, while the electromagnetic
component (which is attributed to the bulk radiation field) is
unconstrained.  It follows that an isolated horizon is
supersymmetric only if it is extremal and nonrotating.  A
remarkable property of these horizons is that the Killing spinor
only has to exist on the horizon itself.  It does not have to exist
off the horizon.  In addition, we find that the limit when the
surface gravity of the horizon goes to zero provides a topological
constraint.  Specifically, the integral of the scalar curvature
of the cross sections of the horizon has to be positive when the
dominant energy condition is satisfied and the cosmological
constant $\Lambda$ is zero or positive, and in particular rules
out the torus topology for supersymmetric isolated horizons
(unless $\Lambda<0$) if and only if the stress-energy tensor
$T_{ab}$ is of the form such that $T_{ab}\ell^{a}n^{b}=0$ for
any two null vectors $\ell$ and $n$ with normalization
$\ell_{a}n^{a}=-1$ on the horizon.

\end{abstract}

\hspace{0.35cm}{\small \textbf{PACS}: 04.70.Bw; 04.20.Fy; 11.30.Pb}



\section{Introduction}

Isolated horizons (IHs) were first introduced as an application of the loop
quantum gravity (LQG) approach to black hole statistical mechanics
\cite{abck,ack,abk}.  It was realized soon after that the framework has a
very rich and elegant classical structure \cite{abf1,abf2}.  An IH is a null
hypersurface at which the intrinsic geometry is held fixed; this generalizes
the notion of an event horizon so that the black hole is an object that is in
local equilibrium with its (possibly) dynamic environment.  Remarkably, the
existence of such a surface is sufficient for the zeroth and first laws of
black-hole mechanics to be satisfied.  However, unlike the older approach
that is based on Killing horizons \cite{wald1,iyewal,jkm}, the first law of
IHs relates quantities that are all defined \emph{at the horizon}.

During the last several years, IHs have been extensively studied.
Calculations involving IHs were first done in terms of the (complex or real)
self-dual connections and $SL(2,\mathbb{C})$ soldering forms
\cite{ack,abk,abf1,abf2}, and later were refined in terms of real Lorentz
connections and tetrads \cite{afk}.  The former approach is better suited
for quantum applications, while the latter for classical mechanics and
geometry.  The framework was extensively studied with inclusion of matter
fields (dilaton, Yang-Mills etc) \cite{ashcor,corsud,acs}, rotation
\cite{abl1}, and a negative cosmological constant in three dimensions
\cite{adw}.  Geometrical issues were extensively studied in
\cite{abl2,plj,lewpaw1}.  Following from earlier work on marginally
trapped surfaces \cite{hayward}, IHs were also extended to nonequilibrium
black holes known as dynamical \cite{ashkri1,ashkri2} and slowly evolving
\cite{boofai1} horizons.

IHs have only recently been extended to higher dimensions, notably for vacuum
spacetimes \cite{lewpaw3,klp}, asymptotically anti-de Sitter (ADS) spacetimes
\cite{apv}, and Einstein-Gauss-Bonnet spacetimes \cite{likboo1}.  Following up
on the latter result, it was shown \cite{liko1} that there is a lower bound on
the GB parameter for which the area-increase law can be violated when two black
holes merge.  The calculation was done for IHs in four dimensions, but presumably
the result holds for IHs in higher dimensions as well (although in the latter case
the topologies are not as severely restricted as they are in four dimensions, even
for Einstein gravity with vanishing cosmological constant \cite{hoy,galsch}).

The purpose of this work is to present a further extension of IHs to supergravity.
In particular, we consider an extension of the isolated horizon framework that
includes charged and rotating black holes in Einstein-Maxwell (EM) theory with an
additional Chern-Simons (CS) term for the electromagnetic field.  The present work
is therefore a quasilocal generalization of the supersymmetric black holes that
were discussed in \cite{gmt}; analogues of these black holes in asymptotically
ADS spacetimes were constructed in \cite{gutrea} and extended to multi-charge
solutions in \cite{klr1}.  Here, we include an arbitrary cosmological constant,
but do not make any assumptions about the boundary at infinity.

We consider the phase space of solutions to the equations of motion for the EMCS
action
\bea
S = \frac{1}{2k_{D}}\int_{\mathcal{M}}d^{D-1}x\sqrt{-g}
    \left[R - 2\Lambda - \frac{1}{4}F^{2} + \beta \epsilon_{ab_{1}\cdots b_{D-1}}
    A^{a}F^{b_{1}b_{2}}\cdots F^{b_{D-2}b_{D-1}}\right] \; .
\label{action1}
\eea
Here, $R$ is the scalar curvature of the Lorentzian manifold $\mathcal{M}$,
$F^{2}=F_{ab}F^{ab}$ ($a,b,\ldots \in\{0,\ldots,D-1\}$) with
$F_{ab}=\partial_{a}A_{b}-\partial_{b}A_{a}$ and $A_{a}$ is the electromagnetic
vector potential.  The constants appearing in the action are the gravitational
coupling constant $k_{D}=8\pi G_{D}$ (with $G_{D}$ the $D$-dimensional Newton
constant), the cosmological constant $\Lambda$ and the CS parameter $\beta$.  The
last term is a CS term for the electromagnetic field that can be added to the
action for odd-dimensional spacetimes.

In Section 2 we rewrite the action (\ref{action1}) in the first-order connection
formulation of general relativity, after which we specify the boundary conditions
that are imposed onto the inner boundary of $\mathcal{M}$; these capture the notion
of a weakly isolated horizon (WIH) that physically corresponds to an isolated black
hole in a surrounding spacetime with (possibly dynamical) fields, imply that the
horizon electric charge is independent of the choice of horizon cross sections,
and leads to the zeroth law of black-hole mechanics.  ``Weak isolation'' is a slightly
less restrictive notion than ``isolation'' in the sense that weak isolation is the
very minimum requirement for the zeroth law to follow from the boundary conditions.

In Section 3 we study the mechanics of the WIHs.  In particular, we show that the
action principle with boundaries is well defined by explicitly showing that the first
variation of the surface term vanishes on the horizon.  We then find an expression for
the symplectic structure by integrating over a spacelike hypersurface the antisymmetrized
second variation of the surface term.  This allows us to find an expression for the
local version of the (equilibrium) first law of black-hole mechanics in dimensions
$D\geq5$.  Summarizing Section 2 and Section 3, we have the following:

\noindent{\bf Result 1.}
\emph{A charged and rotating WIH $\Delta \subset \mathcal{M}$ on the phase space
of solutions of EMCS theory in $D$ dimensions satisfies the zeroth and first laws
of black-hole mechanics.}

Beginning in Section 4 we restrict our study to the stronger notion of (fully) IHs.
The sign of the surface gravity $\kappa_{(\ell)}$ is well defined for such horizons.
The requirement that $\kappa_{(\ell)}\geq0$ therefore allowed us to define a parameter
that provides a constraint on the topology of the IHs.  Specifically, we find that the
integral of the scalar curvature of the cross sections of the horizon (in a spacetime
with nonnegative cosmological constant) have to be strictly positive if the dominant
energy condition is satified or otherwise zero iff the horizon is extremal and
nonrotating with the stress-energy tensor $T_{ab}$ of the form such that
$T_{ab}\ell^{a}n^{b}=0$ for any two null vectors $\ell$ and $n$ with normalization
$\ell_{a}n^{a}=-1$ at the horizon.  In the case of electromagnetic fields with or
without the CS term, the scalar $T_{ab}\ell^{a}n^{b}$ is the square of the electric
flux crossing the horizon.

In Section 5 we specialize to IHs in four and in five dimensions with vanishing
cosmological constant.  With $\beta=0$ the action (\ref{action1}) is the bosonic part
of four-dimensional $N=2$ supergravity, and with $\beta=-2/(3\sqrt{3})$ the action
(\ref{action1}) is the bosonic part of five-dimensional $N=1$ supergravity.  We show that
the existence of a Killing spinor on the IH requires that in both four and in five
dimensions the induced (normal) connection on the horizon has to vanish.  The IH boundary
conditions then imply that the surface gravity and rotation one-form of the horizon are
zero.  This leads to the following:

\noindent{\bf Result 2.}
\emph{An IH of $D=4$ EM theory and of $D=5$ EMCS theory with vanishing cosmological
constant is supersymmetric only if the surface gravity and rotation one-form are zero.}

The angular momentum of an IH in general contains a contribution from the Maxwell fields.
However, this contribution can be shown \cite{abl1} to be a result of the bulk radiation
field integrated to surface terms; the surface term at infinity is zero due to the fall-off
conditions imposed on the Maxwell fields.  Therefore Result 2 implies that a supersymmetric
IH (SIH) is extremal and nonrotating.  The Breckenridge-Myers-Peet-Vafa (BMPV) black hole
\cite{bmpv} is an example of a distorted horizon with arbitrary rotations in the bulk fields
\cite{aepv}; when the angular momentum vanishes this solution reduces to the extremal
Reissner-Nordstr$\ddot{\mbox{o}}$m (RN) solution \cite{myeper} in isotropic coordinates.
The conclusions of Section 4 imply that the only possible horizon topologies for SIHs
are $S^{2}$ in four dimensions while $S^{3}$ and $S^{1}\times S^{2}$ in five dimensions.
The torus topology is a special case that is allowed only if the square of the electric
flux across the horizon is zero.

In Section 6 we conclude with some brief comments about further research that can
be done with the formalism developed here.

\section{Boundary conditions and the zeroth law}

\subsection{First-order action for gravity coupled to electromagnetic fields}

For application to IHs, we work in the first-order formulation with the
``connection-dynamics'' approach.  For details we refer the reader to the
review \cite{ashlew} and references therein.  In this formulation, the configuration
space consists of the triple $(e^{I},A_{\phantom{a}J}^{I},\bm{A})$; the coframe
$e^{I}=e_{a}^{\phantom{a}I}dx^{a}$ determines the metric
\bea
g_{ab} = \eta_{IJ}e_{a}^{\phantom{a}I} \otimes e_{b}^{\phantom{a}J} \, ,
\eea
the gravitational $SO(D-1,1)$ connection
$A_{\phantom{a}J}^{I}=A_{a\phantom{a}J}^{\phantom{a}I}dx^{a}$ determines
the curvature two-form
\bea
\Omega_{\phantom{a}J}^{I}
= dA_{\phantom{a}J}^{I} + A_{\phantom{a}K}^{I} \wedge A_{\phantom{a}J}^{K} \, ,
\eea
and the electromagnetic $U(1)$ connection $\bm{A}$ determines the curvature
\bea
\bm{F} = d\bm{A} \; .
\eea
In this paper, spacetime indices $a,b,\ldots\in\{0,\ldots D-1\}$ are raised
and lowered using the metric tensor $g_{ab}$, while internal Lorentz indices
$I,J,\ldots\in\{0,\ldots,D-1\}$ are raised and lowered using the Minkowski
metric $\eta_{IJ}=\mbox{diag}(-1,1,\ldots,1)$.  The curvature $\Omega$
defines the Riemann tensor $R_{\phantom{a}JKL}^{I}$ (with the convention of
Wald \cite{wald2}) via
\bea
\Omega_{\phantom{a}J}^{I}
= \frac{1}{2}R_{\phantom{a}JKL}^{I}e^{K} \wedge e^{L} \; .
\eea
The Ricci tensor is then $R_{IJ}=R_{\phantom{a}IKJ}^{K}$, and the
Ricci scalar is $R=\eta^{IJ}R_{IJ}$.  The gauge covariant derivative
$\mathscr{D}$ acts on generic fields $\Psi_{IJ}$ such that
\bea
\mathscr{D}\Psi_{\phantom{a}J}^{I}
= d\Psi_{\phantom{a}J}^{I}
  + A_{\phantom{a}K}^{I} \wedge \Psi_{\phantom{a}J}^{K}
  - A_{\phantom{a}J}^{K} \wedge \Psi_{\phantom{a}K}^{I} \; .
\eea
The coframe defines the $(D-m)$-form
\bea
\Sigma_{I_{1}\ldots I_{m}}
= \frac{1}{(D-m)!}\epsilon_{I_{1} \ldots I_{m}I_{m+1} \ldots I_{D}}
e^{I_{m+1}} \wedge \cdots \wedge e^{I_{D}} \, ,
\eea
where the totally antisymmetric Levi-Civita tensor $\epsilon_{I_{1} \ldots I_{D}}$
is related to the spacetime volume element by
\bea
\epsilon_{a_{1} \ldots a_{D}}=\epsilon_{I_{1} \ldots I_{D}}
e_{a_{1}}^{\phantom{a}I_{1}} \cdots e_{a_{D}}^{\phantom{a}I_{D}} \; .
\eea
In this configuration space, the action (\ref{action1}) for the theory
on the manifold $(\mathcal{M},g_{ab})$ (assumed for the moment to have
no boundaries) is given by
\bea
S = \frac{1}{2k_{D}}\int_{\mathcal{M}}\Sigma_{IJ} \wedge \Omega^{IJ}
    - 2\Lambda\bm{\epsilon}
    - \frac{1}{4}\bm{F} \wedge \star \bm{F} + \beta\bm{A} \wedge \bm{F}^{\mathfrak{n}} \; .
\label{action2}
\eea
Here $\bm{\epsilon}=e^{0} \wedge \cdots \wedge e^{D-1}$ is the spacetime
volume element, ``$\star$'' denotes the Hodge dual, and $\mathfrak{n}=(D-1)/2$.

The equations of motion are given by $\delta S=0$, where $\delta$ is
the first variation; i.e. the stationary points of the action.  For this
configuration space the equations of motion are derived from independently
varying the action with respect to the fields $(e,A,\bm{A})$.  To get the
equation of motion for the coframe we note the identity
\bea
\delta\Sigma_{I_{1} \ldots I_{m}}
= \delta e^{M} \wedge \Sigma_{I_{1} \ldots I_{m}M} \; .
\eea
This leads to
\bea
\Sigma_{IJK} \wedge \Omega^{JK} - 2\Lambda\Sigma_{I}
= \mathscr{T}_{I} \, ,
\label{eom1}
\eea
where $\mathscr{T}_{I}$ denotes the electromagnetic stress-energy ($D-1$)-form.
The equation of motion for the connection $A$ is
\bea
\mathscr{D}\Sigma_{IJ} = 0 \, ;
\label{eom2}
\eea
this equation says that the torsion $T^{I}=\mathscr{D}e^{I}$ is zero.  The
equation of motion for the connection $\bm{A}$ is
\bea
d \star \bm{F} + 4(\mathfrak{n}+1)\beta\bm{F}^{\mathfrak{n}} = 0 \; .
\label{eom3}
\eea
The second term in this equation is the contribution due to the CS term in
the action; when this term is turned off, the equation reduces to the standard
Maxwell equation $d \star \bm{F}=0$.  The equations (\ref{eom1}) and (\ref{eom2})
are equivalent to the Einstein equations in the metric formulation, with the
components of $\mathscr{T}_{I}$ identified with the electromagnetic stress-energy
tensor.

\subsection{Boundary conditions}

Let us from here on consider the manifold $(\mathcal{M},g_{ab})$ to contain
boundaries; the conditions that we will impose on the inner boundary will capture
the notion of an isolated black hole that is in local equilibrium with its (possibly)
dynamic surroundings.  For details we refer the reader to \cite{likboo1} and the
references listed there.  We follow the general recipe that was developed in
\cite{afk}, and later extended to rotating IHs in \cite{abl1}.

First, we give some general comments about the structure of the manifold.
Specifically, $(\mathcal{M},g_{ab})$ is a $D$-dimensional Lorentzian manifold with
topology $R\times M$, contains a $(D-1)$-dimensional null surface $\Delta$ as inner
boundary (representing the horizon), and is bounded by $(D-1)$-dimensional spacelike
surfaces $M^{\pm}$ that intersect $\Delta$ in compact $(D-2)$-spaces $\mathbb{S}^{\pm}$
and extend to the boundary at infinity $\mathscr{B}$.
\begin{figure}[h]
\begin{center}
\psfrag{D}{$\Delta$}
\psfrag{Mp}{$M^{+}$}
\psfrag{Mm}{$M^{-}$}
\psfrag{Mi}{$M$}
\psfrag{Sp}{$\mathbb{S}^{+}$}
\psfrag{Sm}{$\mathbb{S}^{-}$}
\psfrag{S}{$\mathbb{S}$}
\psfrag{B}{$\mathscr{B}$}
\psfrag{M}{$\mathcal{M}$}
\includegraphics[width=4.5in]{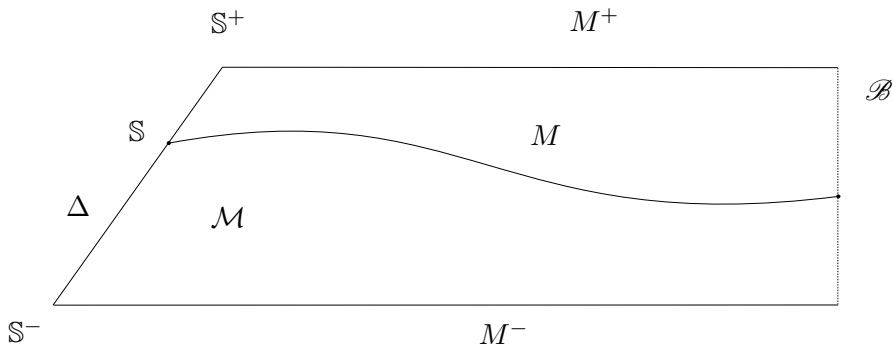}
\caption{The region of the $D$-dimensional spacetime $\mathcal{M}$
being considered has an internal null boundary $\Delta$ representing the
event horizon, and is bounded by $(D-1)$-dimensional spacelike surfaces
$M^{\pm}$ which intersect $\Delta$ in compact $(D-2)$-spaces
$\mathbb{S}^{\pm}$ and extend to the boundary at infinity $\mathscr{B}$.
Here, $M$ is a $(D-1)$-dimensional partial Cauchy surface.}
\end{center}
\end{figure}

The outer boundary $\mathscr{B}$ is some arbitrary $(D-1)$-dimensional surface, and
is loosely referred to as the ``boundary at infinity''.  In this paper, as in
\cite{likboo1}, we consider the purely quasilocal case and neglect any subleties that
are associated with the outer boundary.  Including this contribution in the phase
space amounts to imposing fall-off conditions on the fields for fixed $\Lambda$ (e.g.
asymptotically flat \cite{afk} or asymptotically ADS \cite{apv}) as they approach
$\mathscr{B}$.

Fixing the geometric structures and matter fields on $\Delta$ captures the notion of
a non-expanding horizon \cite{afk}:

\noindent{\bf Definition I.}
\emph{A non-expanding horizon $(\Delta,q_{ab},\ell_{a})$ is a $(D-1)$-dimensional
null hypersurface $\Delta$ with topology $R\times \mathbb{S}^{D-2}$ together with
a degenerate metric $q_{ab}$ of signature $0+\ldots+$ (with $D-2$ nondegenerate
spatial directions) and a null normal $\ell_{a}$ such that: (a) the expansion
$\theta_{(\ell)}$ of $\ell_{a}$ vanishes on $\Delta$; (b) the field equations
hold on $\Delta$; and (c) the stress-energy tensor is such that the vector
$-T_{\phantom{a}b}^{a}\ell^{b}$ is a future-directed and causal vector.}

This is the standard definition for a spacetime that satisfies the Einstein field
equations.  Conditions (a) and (c), together with the Frobenius theorem (which implies
that the rotation tensor $\omega_{ab}$ vanishes on $\Delta$) and the Raychaudhuri
equation, then implies that the shear tensor $\sigma_{ab}$ vanishes on $\Delta$.
The vanishing of all these quantities implies that
\bea
\nabla_{\!\underleftarrow{a}}\ell_{b}\approx\omega_{a}\ell_{b} \, ,
\label{connectionondelta}
\eea
with ``$\approx$'' denoting equality restricted to $\Delta\subset\mathcal{M}$ and
the underarrow indicating pull-back to $\Delta$.  Thus the one-form $\omega$ is the
natural connection (in the normal bundle) induced on the horizon.

A further consequence of the conditions (a) and (c) together with the Raychaudhuri
equation is that the EM stress-energy tensor (which is the usual one for electromagnetism
because variation of the CS term with respect to $g_{ab}$ is zero)
satisfies the constraint \cite{afk}
\bea
T_{ab}\ell^{a}\ell^{b} \approx 0 \; .
\eea
and therefore that
\bea
\underleftarrow{\ell \lrcorner \bm{F}} = 0 \; .
\label{pullback1}
\eea
With the Maxwell-CS equations and the Bianchi identity, it then follows that
\bea
\pounds_{\ell}\underleftarrow{\bm{F}}
\approx \ell \lrcorner \underleftarrow{d\bm{F}}
          + d(\underleftarrow{\ell \lrcorner \bm{F}}) = 0 \; .
\label{liepullback}
\eea
One of the physical consequences of this restriction is that the total electric
charge is independent of the choice of cross sections $\mathbb{S}^{D-2}$.  This
is discussed in detail in \cite{abf2}.

\subsection{Zeroth law}

The ``time-independence'' of $\omega$ and $\underleftarrow{\bm{A}}$ on $\Delta$
captures the notion of a WIH \cite{afk}:

\noindent{\bf Definition II.}
\emph{A WIH $(\Delta,q_{ab},[\ell])$ is a non-expanding horizon $\Delta$ together
with an equivalence class of null normals $[\ell]$ such that
$\pounds_{\ell}\omega_{a}=0$ and $\pounds_{\ell}\underleftarrow{\bm{A}}=0$ for all
$\ell\in[\ell]$ (where $\ell^{\prime}\sim\ell$ if $\ell^{\prime}=c\ell$ for some
constant $c$).}

The above is not a physical restriction.  It is a restriction on the rescaling
freedom of $\ell$ and the gauge freedom of $\bm{A}$.  Now, the surface gravity and
electromagnetic scalar potential are defined respectively as
$\kappa_{(\ell)}=\ell^{a}\omega_{a}$ and $\Phi_{(\ell)}=-\ell^{a}A_{a}$.  In general,
these quantities will change under rescalings of the normal.  However, the conditions
$\pounds_{\ell}\omega_{a}=0$ and $\pounds_{\ell}\bm{A}=0$ are sufficient to ensure
that $d\kappa_{(\ell)}=d(\ell^{a}\omega_{a})=0$ and
$d\Phi_{(\ell)}=d(\ell^{a}\Phi_{a})=0$ \cite{afk}.  The zeroth law therefore follows
from the boundary conditions and is independent of the functional content of the
action (and in particular of the material content of the stress-energy tensor that
describes the coupled matter fields).

\section{Mechanics and the first law}

\subsection{Variation of the boundary term}

Let us now look at the variation of the action (\ref{action2}).  Denoting the
triple $(e,A,\bm{A})$ collectively as a generic field variable $\Psi$, the
first variation gives
\bea
\delta S = \frac{1}{2k_{D}}\int_{\mathcal{M}}E[\Psi]\delta\Psi
           - \frac{1}{2k_{D}}\int_{\partial\mathcal{M}}J[\Psi,\delta\Psi] \; .
\label{first}
\eea
Here $E[\Psi]=0$ symbolically denotes the equations of motion and
\bea
J[\Psi,\delta\Psi] = \Sigma_{IJ} \wedge \delta A^{IJ} - \bm{\Phi} \wedge \delta\bm{A}
\label{surface}
\eea
is the surface term with $(D-2)$-form
\bea
\bm{\Phi} = \star\bm{F} + 4(\mathfrak{n}+1)\beta \bm{A} \wedge \bm{F}^{\mathfrak{n}-1} \; .
\label{boldphi}
\eea
If the integral of $J$ on the boundary $\partial\mathcal{M}$ vanishes then the
action principle is said to be differentiable.  We must show that this is the
case.  Because the fields are held fixed at $M^{\pm}$ and at $\mathscr{B}$, $J$
vanishes there.  Therefore it suffices to show that $J$ vanishes at the inner
boundary $\Delta$.  To show that this is true we need to find an expression for
$J$ in terms of $\Sigma$, $A$ and $\bm{A}$ pulled back to $\Delta$.  As for the
gravitational variables, this is accomplished by employing a higher-dimensional
analogue of the Newman-Penrose (NP) formalism \cite{ppcm}.  In particular, we fix
an internal basis consisting of the (null) pair $(\ell,n)$ and $D-2$ spacelike
vectors $\vartheta_{(i)}$ ($i\in\{2,\ldots,D-1\}$) such that
\bea
e_{0} = \ell \, ,
\quad
e_{1} = n \, ,
\quad
e_{i} = \vartheta_{(i)} \, ,
\label{npbasis}
\eea
together with the conditions
\bea
\ell \cdot n=-1 \, ,
\quad
\ell \cdot \ell = n \cdot n = \ell \cdot \vartheta_{(i)} = n \cdot \vartheta_{(i)} = 0 \, ,
\quad
\vartheta_{(i)} \cdot \vartheta_{(j)} = \delta_{ij} \; .
\label{npconditions}
\eea
In the following we also apply the summation convention over repeated
spacelike indices ($i,j,k$ etc.). As these are Euclidean indices their
position (up or down) will be adjusted according to the dictates of
notational convenience.

To find the pull-back to $\Delta$ of $\Sigma$, we use the
decomposition
\bea
e_{\underleftarrow{a}}^{\phantom{a}I}
\approx -\ell^{I}n_{a} + \vartheta_{(i)}^{\phantom{a}I}\vartheta^{(i)}_{a} \, ,  
\label{pullbackofcoframe}
\eea
whence the $(D-2)$-form
\bea
\underleftarrow{\Sigma}_{IJ} &\approx&
-\frac{1}{(D-3)!}
\epsilon_{IJA_1 \dots A_{D-2}} \ell^{A_1} 
\vartheta^{\phantom{a}A_2}_{(i_1)} \dots \vartheta^{\phantom{a}A_{D-2}}_{(i_{D-3})}
\left(
n \wedge \vartheta^{(i_1)} \wedge \dots \wedge \vartheta^{(i_{D-3})} \right) 
\nonumber \\
& &
+ \frac{1}{(D-2)!} \epsilon_{IJA_1 \dots A_{D-2}}  
\vartheta^{\phantom{a}A_1}_{(i_1)} \dots \vartheta^{\phantom{a}A_{D-2}}_{(i_{D-2})}
\left( \vartheta^{(i_1)} \wedge \dots \wedge \vartheta^{(i_{D-2})} \right) \; .
\label{pullbackofsigmaij}
\eea
To find the pull-back of $A$ we first note that \cite{afk}
\bea
\nabla_{\!\underleftarrow{a}}\ell_{I}
\approx \omega_{a}\ell_{I} \; .
\label{pullback2}
\eea
Then, taking the covariant derivative of $\ell$ acting on internal indices gives
\bea
\nabla_{\! a}\ell_{I} = \partial_{a}\ell_{I} + A_{aIJ}\ell^{J} \, ,
\eea
with $\partial$ representing a flat derivative operator that is compatible with
the internal coframe on $\Delta$.  Thus $\partial_{a}\ell_{I}\approx0$ and
\bea
\nabla_{\!\underleftarrow{a}}\ell_{I}\approx A_{\underleftarrow{a}IJ}\ell^{J} \; .
\eea
Putting this together with (\ref{pullback2}) we have that
\bea
A_{\underleftarrow{a}IJ}\ell^{J}\approx\omega_{a}\ell_{I} \, ,
\eea
and this implies that the pull-back of $A$ to the horizon is of the form
\bea
A_{\underleftarrow{a}}^{\phantom{a}IJ}
\approx -2\ell^{[I}n^{J]}\omega_{a}
+ a_{a}^{(i)}\ell^{[I}\vartheta_{(i)}^{\phantom{a}J]}
+ b_a^{(ij)}\vartheta_{(i)}^{\phantom{a}[I}\vartheta_{(j)}^{\phantom{a}J]} \, ,
\label{pullbackofa}
\eea
where the $a_{a}^{(i)}$ and $b_{a}^{(ij)}$ are one-forms in the cotangent
space $T^{*}(\Delta)$.  It follows that the variation of (\ref{pullbackofa})
is
\bea
\delta A_{\underleftarrow{a}}^{\phantom{a}IJ}
\approx -2\ell^{[I}n^{J]}\delta \omega_{a}
+ \delta a_{a}^{(i)}\ell^{[I}\vartheta_{(i)}^{\phantom{a}J]}
+ \delta b_a^{(ij)} \vartheta_{(i)}^{\phantom{a}[I}\vartheta_{(j)}^{J]} \; .
\label{variationofpullbackofa}
\eea
Finally, by direct calculation, it can be shown that the gravitational part
$J_{\rm Grav}$ of the surface term (\ref{surface}) reduces to
\bea
J_{\rm Grav}[\Psi,\delta\Psi]
\approx \bm{\tilde{\epsilon}} \wedge \delta \omega \; .
\label{simplifiedpullbackofcurrent}
\eea
Here, $\bm{\tilde{\epsilon}}=\vartheta^{(1)} \wedge \dots \wedge \vartheta^{(D-2)}$
is the area element of the cross sections $\mathbb{S}^{D-2}$ of the horizon.

Now we make use of the fact that, because $\ell$ is normal to the surface, its variation
will also be normal to the surface.  That is, $\delta\ell\propto\ell$ for some $\ell$
fixed in $[\ell]$.  This together with $\pounds_{\ell}\omega=0$ then implies that
$\pounds_{\ell}\delta\omega=0$.  However, $\omega$ is held fixed on $M^{\pm}$ which
means that $\delta\omega=0$ on the initial and final cross-sections of $\Delta$
(i.e. on $M^{-}\cap\Delta$ and on $M^{+}\cap\Delta$), and because $\delta\omega$ is
Lie dragged on $\Delta$ it follows that $J_{\rm Grav}\approx0$.  The same argument
also holds for the electromagnetic part $J_{\rm EM}$ of the surface term (\ref{surface}).
In particular, because the electromagnetic field is in a gauge adapted to the horizon,
$\pounds_{\ell}\underleftarrow{\bm{A}}=0$, and with $\delta\ell\propto\ell$ we also have
that $\pounds_{\ell}\delta\underleftarrow{\bm{A}}=0$.  This is sufficient to show that
$J_{\rm EM}\approx0$ as well.  Therefore the surface term $J|_{\partial\mathcal{M}}=0$
for the Einstein-Maxwell theory with electromagnetic CS term, and we conclude that the
equations of motion $E[\Psi]=0$ follow from the action principle $\delta S=0$.

\subsection{Covariant phase space}

The derivation of the first law involves two steps.  First we need to find the symplectic
structure on the covariant phase space $\bm{\Gamma}$ consisting of solutions $(e,A,\bm{A})$
to the field equations (\ref{eom1}), (\ref{eom2}) and (\ref{eom3}) on $\mathcal{M}$.  Once
we have a suitable (closed and conserved) symplectic two-form, we then need to specify an
evolution vector field $\xi^{a}$.  In this section we derive the symplectic two-form.  In
the next section we will specify the evolution vector field which will also serve to
introduce an appropriate notion of horizon angular momentum.

The antisymmetrized second variation of the surface term gives the symplectic current,
and integrating over a spacelike hypersurface $M$ gives the symplectic structure
$\bm{\Omega}\equiv\bm{\Omega}(\delta_{1},\delta_{2})$ (with the choice of $M$ being
arbitrary).  Following \cite{afk}, we find that the second variation of the surface term
(\ref{surface}) gives
\bea
J[\Psi,\delta_{1}\Psi,\delta_{2}\Psi]
= \delta_{1}\Sigma_{IJ} \wedge \delta_{2}A^{IJ}
  - \delta_{2}\Sigma_{IJ} \wedge \delta_{1}A^{IJ}
  - \delta_{1}\bm{\Phi} \wedge \delta_{2}\bm{A}
  - \delta_{2}\bm{\Phi} \wedge \delta_{1}\bm{A} \; .
\eea
Whence integrating over $M$ defines the \emph{bulk} symplectic structure
\bea
\bm{\Omega}_{\rm bulk}
= \frac{1}{2k_{D}}\int_{M}
  \left[\delta_{1}\Sigma_{IJ} \wedge \delta_{2}A^{IJ}
  - \delta_{2}\Sigma_{IJ} \wedge \delta_{1}A^{IJ}
  - \delta_{1}\bm{\Phi} \wedge \delta_{2}\bm{A}
  + \delta_{2}\bm{\Phi} \wedge \delta_{1}\bm{A}\right] \; .
\label{bulksymplectic}
\eea
We also need to find the pull-back of $J$ to $\Delta$ and add the integral of this
term to $\bm{\Omega}_{\rm bulk}$ so that the resulting symplectic structure on
$\bm{\Gamma}$ is conserved.  If we define potentials $\psi$ and $\chi$ for the
surface gravity $\kappa_{(\ell)}$ and electric potential $\Phi_{(\ell)}$ such that
\bea
\pounds_{\ell}\psi \approx \ell \lrcorner \omega = \kappa_{(\ell)}
\quad
\mbox{and}
\quad
\pounds_{\ell}\chi \approx \ell \lrcorner \bm{A} = -\Phi_{(\ell)} \, ,
\eea
then the pullback to $\Delta$ of the symplectic structure will be a total
derivative; using the Stokes theorem this term becomes an integral over the cross
sections $\mathbb{S}^{D-2}$ of $\Delta$.  Hence the full symplectic structure is given by
\bea
\bm{\Omega}
&=& \frac{1}{2k_{D}}\int_{M}
    \left[\delta_{1}\Sigma_{IJ} \wedge \delta_{2}A^{IJ}
    - \delta_{2}\Sigma_{IJ} \wedge \delta_{1}A^{IJ}
    - \delta_{1}\bm{\Phi} \wedge \delta_{2}\bm{A}
    + \delta_{2}\bm{\Phi} \wedge \delta_{1}\bm{A}\right]\nonumber\\
& &
    + \frac{1}{k_{D}}\oint_{\mathbb{S}^{D-2}}
    \left[\delta_{1}\bm{\tilde{\epsilon}} \wedge \delta_{2}\psi
    - \delta_{2}\bm{\tilde{\epsilon}} \wedge \delta_{1}\psi
    + \delta_{1}\bm{\Phi} \wedge \delta_{2}\chi
    - \delta_{2}\bm{\Phi} \wedge \delta_{1}\chi\right] \; .
\label{fullsymplectic}
\eea

\subsection{Angular momentum and the first law}

In $D$ dimensions, there are $\tilde{\mathfrak{n}}=\lfloor (D-1)/2 \rfloor$
rotation parameters given by the Casimirs of the rotation group $SO(D-1)$.  Here,
``$\lfloor (D-1)/2 \rfloor$'' denotes the integer value of $\mathfrak{n}$.  For a
multidimensional WIH rotating with angular velocities $\Omega_{\iota}$
($\iota=1,\ldots,\tilde{\mathfrak{n}}$), a suitable evolution vector field on the
covariant phase space is given by \cite{abl1,apv}
\bea
\xi^{a} = k\ell^{a} + \sum_{\iota=1}^{\tilde{\mathfrak{n}}}\Omega_{\iota}\phi_{\iota}^{a} \; .
\eea
Here, $k$ is a constant on $\Delta$ and $\phi_{\iota}^{a}$ are spacelike rotational
vector fields that satisfy
\bea
\pounds_{\phi}q_{ab}=0 \, ,
\quad
\pounds_{\phi}\ell_{a}=0 \, ,
\quad
\pounds_{\phi}\omega_{a}=0 \, ,
\quad
\pounds_{\phi}\underleftarrow{\bm{A}}=0 \, ,
\quad
\pounds_{\phi}\underleftarrow{\bm{F}}=0 \; .
\eea
The vector field $\xi$ is similar to the linear combination
$\zeta=t+\sum_{\iota}\Omega_{\iota}m_{\iota}$ (with $t$ a timelike Killing vector and
$m_{\iota}$ spacelike Killing vectors) for globally stationary spacetimes which on a
Killing horizon is null.  By contrast, we note that $\xi$ is null on $\Delta$ only
when all angular momenta are zero; in general $\xi$ is \emph{spacelike}.

Moving on, the first law now follows directly from evaluating the symplectic
structure at $(\delta,\delta_{\xi})$.  See e.g. \cite{apv} for details.  This
gives two surface terms: one at infinity (which is identified with the ADM
energy), and one at the horizon.  We find that the surface term at the horizon is
given by
\bea
\bm{\Omega}|_{\Delta}
= \frac{\kappa_{(k\ell)}}{k_{D}}\delta\oint_{\mathbb{S}^{D-2}}\bm{\tilde{\epsilon}}
    + \frac{\Phi_{(k\ell)}}{k_{D}}\delta\oint_{\mathbb{S}^{D-2}}\bm{\Phi}
    + \sum_{\iota=1}^{\tilde{\mathfrak{n}}}\frac{\Omega_{\iota}}{k_{D}}
    \delta\oint_{\mathbb{S}^{D-2}}\left[(\phi_{\iota} \lrcorner \omega)\bm{\tilde{\epsilon}}
    + (\phi_{\iota} \lrcorner \bm{A})\bm{\Phi}\right] \, ,
\label{evaluation}
\eea
where we used $\kappa_{(k\ell)}=\pounds_{k\ell}\psi=k\ell\lrcorner\omega$ and
$\Phi_{(k\ell)}=\pounds_{k\ell}\chi=k\ell\lrcorner\omega$.  These potentials are constant
for any given horizon, but in general vary across the phase space from one point to another.
This implies that (\ref{evaluation}) is \emph{not} in general a total variation.  However,
an appropriate normalization of the vector $k\ell^{a}$ can be chosen so that
(\ref{evaluation}) is a total variation.  The necessary and sufficient condition for the
system to be Hamiltonian is that there exists a function $\mathcal{E}$ such that
$\bm{\Omega}|_{\Delta}(\delta,\delta_{\xi})=\delta \mathcal{E}$.  Given that such a
function does exist, we find that
\bea
\delta\mathcal{E} = \frac{\kappa_{(t)}}{k_{D}}\delta\oint_{\mathbb{S}^{D-2}}\bm{\tilde{\epsilon}}
                    + \frac{\Phi_{(k\ell)}}{k_{D}}\delta\oint_{\mathbb{S}^{D-2}}\bm{\Phi}
                    + \sum_{\iota=1}^{\tilde{\mathfrak{n}}}\frac{\Omega_{\iota}}{k_{D}}
                    \delta\oint_{\mathbb{S}^{D-2}}\left[(\phi_{\iota} \lrcorner
                    \omega)\bm{\tilde{\epsilon}}
                    + (\phi_{\iota} \lrcorner \bm{A})\bm{\Phi}\right] \; .
\label{firstlaw}
\eea
This is the first law of black-hole mechanics.  For a quasi-static process, the standard
form of the first law of thermodynamics is
\bea
\delta \mathcal{E} = T_{(k\ell)}\delta \mathcal{S} + \Phi_{(k\ell)}\delta \mathcal{Q}
                     + \sum_{\iota=1}^{\tilde{\mathfrak{n}}}
                       \Omega_{\iota}\delta \mathcal{J}_{\iota} \, ;
\eea
comparing this to (\ref{firstlaw}), and with the identification of the Hawking temperature
$T_{(k\ell)}=\kappa_{(k\ell)}/(2\pi)$, we find that the entropy, electric charge and angular
momenta of the isolated horizon are:
\bea
\mathcal{S}     &=& \frac{1}{4G_{D}}\oint_{\mathbb{S}^{D-2}}\bm{\tilde{\epsilon}}
\label{entropy}\\
\mathcal{Q}     &=& \frac{1}{8\pi G_{D}}\oint_{\mathbb{S}^{D-2}}
                    \star\bm{F}+4(\mathfrak{n}+1)\beta\bm{A} \wedge \bm{F}^{\mathfrak{n}-1}
\label{charge}\\
\mathcal{J}_{\iota} &=& \frac{1}{8\pi G_{D}}\oint_{\mathbb{S}^{D-2}}
                    \left[(\phi_{\iota} \lrcorner \omega)\bm{\tilde{\epsilon}}
                    + (\phi_{\iota} \lrcorner \bm{A})\bm{\Phi}\right] \; .
\label{angularmomentum}
\eea
Therefore WIHs in $D$-dimensional EMCS theory satisfy the first law (and the zeroth law)
of black-hole mechanics.  This is in agreement with \cite{gmt}, but with a very important
difference.  Here, all the quantities appearing in the first law are defined at the horizon;
no reference was made to the boundary at infinity.

\noindent{\bf Remark.}
The expression (\ref{angularmomentum}) implies that the horizon angular momentum
contains contributions from both gravitational \emph{and} electromagnetic fields,
here referred to as $\mathcal{J}_{\rm Grav}$ and $\mathcal{J}_{\rm EM}$.  This is
in contrast to the standard angular momentum expressions at infinity, such as the
Komar expression.  One can show \cite{abl1,apv} that $\mathcal{J}_{\rm Grav}$ is
equivalent to the (quasilocal) Komar integral
\bea
\mathcal{J}_{\rm K} = -\frac{1}{8\pi G_{D}}\oint_{\mathbb{S}^{D-2}}\star d\phi \, ,
\eea
and this matches the expression for Killing horizons at infinity.

It would appear that if $\omega=0$ then there is still a nonzero contribution
to (\ref{angularmomentum}) from $\mathcal{J}_{\rm EM}$.  However, it can be shown
\cite{abl1,apv} that if $\phi$ is the restriction to $\Delta$ of a \emph{global}
rotational Killing field $\varphi$ contained in $\mathcal{M}$, then
$\mathcal{J}_{\rm EM}$ is actually the angular momentum of the electromagnetic radiation
in the bulk.  What happens is that the bulk integral can be written as the sum of a
surface term at $\Delta$ and a surface term at $\mathscr{B}$, and the latter vanishes
due to the fall-off conditions on the fields.  Therefore we say that a nonrotating WIH
is one for which $\omega=0$.

\section{A topological constraint from extremality}

One of the properties of an extremal black hole is that its surface gravity is zero.
Another property is that its horizons are degenerate: the inner and outer horizons
coincide.  As a result, an extremal black hole is one for which there are no trapped
surfaces ``just inside'' the horizon.  This property was recently used \cite{boofai2}
to define an extremality condition for quasilocal horizons.  We note here the evolution
equation for the expansion of the null normal $n^{a}$ \cite{boofai2}:
\bea
\pounds_{\ell}\theta_{(n)} + \kappa \theta_{(n)} + \frac{1}{2}\mathcal{R}
= d_{a}\tilde{\omega}^{a} + \|\tilde{\omega}\|^{2} + (T_{ab}
  - \Lambda g_{ab})\ell^{a}n^{b} \; .
\label{nevolution}
\eea
Here, $\mathcal{R}$ is the scalar curvature of $\mathbb{S}^{D-2}$, $d_{a}$ is the
covariant derivative operator that is compatible with the metric
$\tilde{q}_{ab}=g_{ab}+\ell_{a}n_{b}+\ell_{b}n_{a}$, and
$\|\tilde{\omega}\|^{2}=\tilde{\omega}_{a}\tilde{\omega}^{a}$ where
\bea
\tilde{\omega}_{a} = \tilde{q}_{a}^{\phantom{a}b}\omega_{b}
                   = \omega_{a} + \kappa_{(\ell)}n_{a}
\label{amoneform}
\eea
is the projection of $\omega$ onto $\mathbb{S}^{D-2}$.  $\tilde{\omega}$ is referred
to as the rotation one-form.

Our desire is to apply the expression (\ref{nevolution}) to \emph{black holes},
and in order to do this we need to impose some restrictions on the WIHs.
Henceforth we will restrict our attention to (fully) IHs.  These are WIHs for
which there is a scaling of the null normals for which the commutator
$[\pounds_{\ell},\mathcal{D}]=0$ (with $\mathcal{D}$ the intrinsic covariant
derivative on the horizon).  Physically these IHs have time-invariant extrinsic
(as well as intrinsic) geometries and, up to a free constant, a preferred scaling
of the null normals.  The condition for full isolation, however, still does not
guarentee that we are dealing with black holes.  There are many examples of IHs
which do not describe black holes.  An example is a vanishing scalar invariant
(VSI) spacetime for which it has been shown that no trapped surfaces exist
\cite{senovilla,plj}.  Thus we restrict our attention to IHs for which there
do exist trapped surfaces, in which case then $\theta_{(n)}<0$.  Then the sign
of the surface gravity is well defined and a horizon is sub-extremal if
$\kappa>0$ and extremal (with degenerate horizons) if $\kappa=0$.  Further,
$\pounds_{\ell}\theta_{(n)}=0$ and combining this with the fact that the inward
expansion $\theta_{(n)}$ should always be less than zero, an integration of
(\ref{nevolution}) gives
\bea
\eta \equiv \oint_{\mathbb{S}^{D-2}}\bm{\tilde{\epsilon}}(T_{ab}\ell^{a}n^{b}
+ \Lambda + \|\tilde{\omega}\|^{2})
- \frac{1}{2}\oint_{\mathbb{S}^{D-2}}\bm{\tilde{\epsilon}}\mathcal{R} \leq 0 \; .
\label{characterization}
\eea
(Here we used $-\Lambda g_{ab}\ell^{a}n^{b}=\Lambda$ to simplify the cosmological term,
and the fact that $\oint_{\mathbb{S}^{D-2}}\bm{\tilde{\epsilon}}d_{a}\tilde{\omega}^{a}=0$.)
This inequality provides an alternative characterization of extremal IHs: if $\eta<0$
($\kappa>0$ and $\theta_{(n)}<0$) then $\Delta$ is nonextremal, and if $\eta=0$
($\kappa=0$) then $\Delta$ is extremal.  However, this inequality also provides a
topological constraint on the cross sections of $\Delta$.  To see this, rewrite
(\ref{characterization}) so that
\bea
\oint_{\mathbb{S}^{D-2}}\bm{\tilde{\epsilon}}(\mathcal{R} - 2\Lambda)
\geq 2\oint_{\mathbb{S}^{D-2}}\bm{\tilde{\epsilon}}(T_{ab}\ell^{a}n^{b}
+ \|\tilde{\omega}\|^{2}) \; .
\label{curvaturecondition}
\eea
Now, observe that the dominant energy condition requires that
$T_{ab}\ell^{a}n^{b}\geq0$.  In addition, $\|\tilde{\omega}\|^{2}$ is manifestly
non-negative.  The inequality (\ref{curvaturecondition}) therefore restricts the
topology of the cross sections of the horizon.  The condition (\ref{curvaturecondition})
is the same as the one that was found in four dimensions for marginally trapped surfaces
\cite{hayward}, nonexpanding horizons \cite{plj} and dynamical horizons
\cite{ashkri2,boofai3}.

For nonextremal horizons, $\eta<0$, and the constraint (\ref{curvaturecondition})
splits into two possibilities, depending on the nature of the cosmological constant:
\begin{itemize}

\item
$\Lambda \geq 0$.  The integral of the scalar curvature is strictly positive.
In four dimensions the GB theorem says that
$\oint_{\mathbb{S}^{2}}\tilde{\epsilon}\mathcal{R}=8\pi(1-g)$, with $g$ the genus
of the surface $\mathbb{S}^{2}$.  In this case $\eta<0$ implies that $g=0$ and hence
the only possibility is that the cross sections are two-spheres $S^{2}$.  In five
dimensions $\eta<0$ implies that the cross sections are of positive Yamabe type;
this implies that topologically $\mathbb{S}^{3}$ can only be a finite connected
sum of the three-sphere $S^{3}$ or of the ring $S^{1}\times S^{2}$
\cite{galsch,galloway}.  Both these topologies have been realized and the
corresponding solutions, for example the Myers-Perry black hole \cite{myeper}
and the Emparan-Reall black ring \cite{emprea1}, are well known.

\item
$\Lambda < 0$.  The integral of the scalar curvature can have either sign, or even
vanish, and the inequality will always be satisfied.  The only restriction is that
\bea
\oint_{\mathbb{S}^{D-2}}\bm{\tilde{\epsilon}}(\mathcal{R} + 2|\Lambda|)
\geq 2\oint_{\mathbb{S}^{D-2}}\bm{\tilde{\epsilon}}(T_{ab}\ell^{a}n^{b}
+ \|\tilde{\omega}\|^{2}) \; .
\eea
There is no constraint on the topology of $\mathbb{S}^{D-2}$.  Owing to this special
property, and to the realization that black holes can be constructed with suitable
identifications of points in ADS spacetime, many such black holes have been found with
exotic topologies in $D\geq3$ dimensions.  See e.g. \cite{bhtz,abhp,abbhp,banados,bgm}.

\end{itemize}
For extremal horizons, $\eta=0$, and the constraint (\ref{curvaturecondition})
becomes an equality.  In this case the same restrictions apply to
$\oint_{\mathbb{S}^{D-2}}\bm{\tilde{\epsilon}}\mathcal{R}$ as for nonextremal horizons.
However, there is also a special case that occurs:
\bea
\oint_{\mathbb{S}^{D-2}}\bm{\tilde{\epsilon}}\mathcal{R} = 0
\eea
for an extremal and nonrotating ($\omega=\tilde{\omega}=0$) horizon when the scalar
$T_{ab}\ell^{a}n^{b}$ vanishes on the horizon.  This case corresponds to the torus
topology $T^{D-2}$.

\section{Supersymmetric isolated horizons}

\subsection{Killing spinors in $D=4$}

So far in this paper we have discussed physical aspects of WIHs and IHs in arbitrary
dimensions.  However, the main purpose here is to find the conditions for supersymmetry.
Thus we will now examine the special cases of IHs in $D=4$ and $D=5$ dimensions.  From
here on we will also restrict our covariant phase space to solutions for which $\Lambda=0$.

We will first consider the four-dimensional action with $\beta=0$.  This is just
EM theory, which can be embedded into $N=2$ supergravity; the (extremal) RN black
hole, which is a solution to the EM theory, is also a solution to the $N=2$
supergravity with the fermion fields set to zero.  As was shown in \cite{gibhul},
the condition for a supersymmetric black hole in four dimensions to have positive
energy is that $M_{\infty}=|Q_{\infty}|$ which is the extremality condition for
the RN black hole relating the mass $M_{\infty}$ and charge $Q_{\infty}$ at
infinity.  This is also the saturated Bogomol'ny-Prasad-Sommerfeld (BPS) inequality.

In general, solutions to the supergravity equations of motion have to be invariant
under the supersymmetry transformations of the fields.  Black holes in particular
are solutions to the bosonic equations of motion, which means that the fermion
fields vanish.  This implies that the transformations of the bosonic fields $e$
and $\bm{A}$ vanish when evaluated on the solutions, and therefore the only
condition for supersymmetry comes from the fermion variation: the necessary and
sufficient condition for supersymmetry with vanishing fermion fields is that there
exists a Killing spinor $\psi_{AA^{\prime}}=(\alpha_{A},\beta_{A^{\prime}})$
($A,B,\ldots\in\{1,2\}$ and $A^{\prime},B^{\prime},\ldots\in\{1,2\}$) such that
\cite{tod1,tod2}
\bea
\nabla_{AA^{\prime}}\alpha_{B} + \sqrt{2}\phi_{AB}\beta_{A^{\prime}} &=& 0
\label{cond1}\\
\nabla_{AA^{\prime}}\beta_{B^{\prime}} - \sqrt{2}\bar{\phi}_{A^{\prime}B^{\prime}}
\alpha_{A} &=& 0 \; .
\label{cond2}
\eea
Here, $\phi_{AB}$ is the Maxwell spinor and $\bar{\phi}_{A^{\prime}B^{\prime}}$ is its
complex conjugate.  These are related to the field strength via
\bea
\bm{F} = \phi_{AB}\epsilon_{A^{\prime}B^{\prime}}
         + \bar{\phi}_{A^{\prime}B^{\prime}}\epsilon_{AB} \, ;
\eea
the spinor symplectic structure is defined such that $\epsilon^{12}=-\epsilon^{21}=1$.
Using the spinors $\alpha$ and $\beta$ we can define the following set of null vectors:
\bea
\ell_{AA^{\prime}} = \alpha_{A}\bar{\alpha}_{A^{\prime}} \, ,
\quad
n_{AA^{\prime}} = \bar{\beta}_{A}\beta_{A^{\prime}} \, ,
\quad
\vartheta_{AA^{\prime}} = \alpha_{A}\beta_{A^{\prime}} \, ,
\quad
\bar{\vartheta}_{AA^{\prime}} = \bar{\beta}_{A}\bar{\alpha}_{A^{\prime}} \; .
\label{spinvectors}
\eea
It can be shown that the vector
\bea
K_{AA^{\prime}} \equiv \ell_{AA^{\prime}} + n_{AA^{\prime}}
\eea
is a Killing vector; the norm of this vector is given by
\bea
\| K\| = 2V\bar{V} \, ,
\label{tnvector1}
\eea
where we defined the scalar $V=\alpha_{A}\bar{\beta}^{A}$.  It follows that
$K_{AA^{\prime}}$ can be either timelike (referred to as nondegenerate) when
$V\neq0$ or null (referred to as degenerate) when $V=0$.

For IHs, the case of interest is the one for which the Killing spinor is null.
This is a particularly special case because $V=\alpha_{A}\bar{\beta}^{A}=0$
implies that
\bea
\bar{\beta}^{A} = \mathcal{K}\alpha^{A}
\eea
for some function $\mathcal{K}$.  Putting this into the conditions (\ref{cond1})
and (\ref{cond2}) allows one to find an expression for the covariant derivative
in terms of the spinors \cite{tod1,tod2}:
\bea
\nabla_{AA^{\prime}}\alpha_{B}
= -\sqrt{2}\bar{\mathcal{K}}\phi\alpha_{A}\alpha_{B}\bar{\alpha}_{A^{\prime}} \; .
\eea
Here, $\phi$ is a function defined by $\phi_{AB}=\phi\alpha_{A}\alpha_{B}$.  Let
us use this form of the covariant derivative to find the covariant derivative of the
null normal $\ell$ of an IH.  We find that
\bea
\nabla_{a}\ell_{b} = -\sqrt{2}(\bar{\mathcal{K}}\phi + \mathcal{K}\bar{\phi})
                     \ell_{a}\ell_{b} \; .
\eea
This immediately implies that
\bea
\nabla_{\!\underleftarrow{a}}\ell_{b}\approx0 \, ,
\eea
and with (\ref{connectionondelta}) it follows that $\omega=0$.  Therefore the
existence of a Killing spinor on an IH implies that the connection one-form has
to be zero by the boundary conditions.  We emphasize, however, that in this
calculation the Killing spinor only has to exist on the horizon itself.  It
does not have to exist off the horizon.  We will discuss the implications of
this constraint in Subsection 5.3, after we look at the five-dimensional EMCS
theory.

\subsection{Killing spinors in $D=5$}

We will now consider the five-dimensional action.  With $D=5$ and
$\beta=-2/(3\sqrt{3})$ the action (\ref{action2}) is the bosonic sector of $N=1$
supergravity.  As in the four-dimensional EM theory, solutions to the bosonic
equations of motion require the existence of a Killing spinor to ensure that
supersymmetry is preserved.  For black holes, the positive energy theorem together
with this requirement imply that the mass and charge are constrained such that
$M_{\infty}=(\sqrt{3}/2)|Q_{\infty}|$ \cite{gkltt}.  As can be verified, this
equality is satisfied by the (5D) extremal RN black hole \cite{myeper}, the BMPV
black hole \cite{bmpv} and the Elvang-Emparan-Mateos-Reall (EEMR) black ring
\cite{eemr}.

The strategy for finding supersymmetric solutions to the bosonic equations of motion
based on Killing spinors is essentially the same in five dimensions as it is in four
dimensions.  Using the field equations, one determines the constraints on various
bosonic bilinears that are constructed from the Killing spinor, which are then solved
to derive the spacetime metrics and their associated Maxwell fields.  This is the
strategy that was pioneered by Tod for supergravity in four dimensions, and later
extended to five dimensions by Gauntlett \emph{et al} \cite{gghpr}.

The situation in $D\geq5$ dimensions is more complicated from an algebraic point
of view because, unlike the case of supergravity in four dimensions, we cannot use
the simple Newman-Penrose basis as we did in the preceeding subsection.  An additional
problem arises specifically in five dimensions -- spinors satisfying certain reality
conditions cannot be consistently defined unless they come in pairs and are equipped
with a symplectic structure.  For details we refer the interested reader to the
excellent Les Houches lectures by Van Nieuwenhuizen \cite{vannieuwenhuizen}.

In what follows, we shall employ the conventions of \cite{gghpr} and consider (commuting)
symplectic Majorana spinors $\epsilon^{\alpha}$ ($\alpha,\beta,\ldots\in\{1,2\}$; we keep
in mind that unlike the spinor indices $A,A^{\prime}$ in the preceeding subsection, these
spinor indices denote spinors rather than the components of spinors) together with a set
of gamma matrices $\Gamma^{I}$ that satisfy the anticommutation rule
\bea
\Gamma^{I}\Gamma^{J} + \Gamma^{J}\Gamma^{I} = 2\eta^{IJ}
\eea
and the antisymmetry product
\bea
\Gamma_{IJKLM} = \epsilon_{IJKLM} \; .
\eea
In general, $\Gamma_{I_{1}\ldots I_{D}}$ denotes the antisymmetrized product
of $D$ gamma matrices.

The spinors $\epsilon^{\alpha}$ satisfy the reality condition
\bea
\bar{\epsilon}^{\alpha} \equiv (\epsilon^{\alpha})^{\dagger}\Gamma_{0}
= (\epsilon^{\alpha})^{T}\mathcal{C} \, ;
\eea
$\dagger$ denotes Hermitian conjugation, $T$ denotes matrix transpose,
and $\mathcal{C}$ is the charge conjugation operator satisfying
\bea
\mathcal{C}(\Gamma_{I})^{T}\mathcal{C}^{-1} = \Gamma_{I} \; .
\eea
Spinor indices are raised and lowered using the symplectic structure
$\epsilon_{\alpha\beta}$ which is defined such that $\epsilon_{12}=\epsilon^{12}=+1$.

Just as for bosonic fields in four dimensions, the necessary and sufficient
condition for supersymmetry with vanishing fermion fields is that there
exists a Killing spinor.  In the case of five-dimensional supergravity,
$\epsilon^{\alpha}$ are Killing spinors iff
\bea
\left[\nabla_{I} + \frac{1}{4\sqrt{3}}(\Gamma_{I}^{\phantom{a}JK}
- 4\delta_{I}^{\phantom{a}J}\Gamma^{K})F_{JK}\right]\epsilon^{\alpha} = 0 \; .
\label{cond3}
\eea
Using $\epsilon^{\alpha}$ we can construct bosonic bilinears $f$, $V^{I}$ and
$\Phi^{\alpha\beta}=\Phi^{(\alpha\beta)}$ such that
\bea
f\epsilon^{\alpha\beta} = \bar{\epsilon}^{\alpha}\epsilon^{\beta} \, ,
\quad
V^{I}\epsilon^{\alpha\beta} = \bar{\epsilon}^{\alpha}\Gamma^{I}\epsilon^{\beta} \, ,
\quad
\Phi^{\alpha\beta IJ} = \bar{\epsilon}^{\alpha}\Gamma^{IJ}\epsilon^{\beta} \; .
\eea
We can find various algebraic relations between these bilinears by using the
Fierz rearrangement identity for the product of four spinors
$\bar{\epsilon}_{1},\epsilon_{2},\bar{\epsilon}_{3}$ and $\epsilon_{4}$, given by
\bea
\bar{\epsilon}_{1}\epsilon_{2}\bar{\epsilon}_{3}\epsilon_{4}
= \frac{1}{4}\left(\bar{\epsilon}_{1}\epsilon_{4}\bar{\epsilon}_{3}\epsilon_{2}
  + \bar{\epsilon}_{1}\Gamma_{I}\epsilon_{4}\bar{\epsilon}_{3}\Gamma^{I}\epsilon_{2}
  - \frac{1}{2}\bar{\epsilon}_{1}\Gamma_{IJ}\epsilon_{4}\bar{\epsilon}_{3}\Gamma^{IJ}
  \epsilon_{2}\right) \; .
\eea
Of particular interest for our purposes is the identity with
$\bar{\epsilon}_{1}=\bar{\epsilon}^{\alpha}$, $\epsilon_{2}=\epsilon^{\delta}$,
$\bar{\epsilon}_{3}=\bar{\epsilon}^{\gamma}$ and $\epsilon_{4}=\epsilon^{\beta}$.
Then we find that
\bea
\epsilon^{\alpha\delta}\epsilon^{\gamma\beta}f^{2}
= \frac{1}{4}\epsilon^{\alpha\beta}\epsilon^{\gamma\delta}(f^{2} + V_{I}V^{I})
  - \frac{1}{8}\Phi_{\phantom{aa}IJ}^{\alpha\beta}\Phi^{\gamma\delta IJ} \, ,
\eea
and contracting both sides of this equation with
$\epsilon_{\alpha\beta}\epsilon_{\gamma\delta}$ leads to
\bea
V_{I}V^{I} = f^{2} \; .
\label{tnvector2}
\eea
This implies that the vector $V^{I}$ is either timelike (if $f\neq0$) or null (if $f=0$).

As for IHs in four dimensions, we are interested in the null case.  Using the Killing spinor
equation (\ref{cond3}) and employing the Fierz identity, it can be shown that the covariant
derivative of $V^{I}$ is given by \cite{gghpr}
\bea
\nabla_{I}V_{J} = \frac{1}{2\sqrt{3}}\epsilon_{IJKLM}F^{KL}V^{M} \; .
\label{cond4}
\eea
Using this, we conclude that an IH of five-dimensional EMCS theory, equipped with a null
normal $\ell$, will be supersymmetric if
\bea
\nabla_{a}\ell_{b} &=& e_{a}^{\phantom{a}I}\nabla_{I}(e_{b}^{\phantom{a}J}\ell_{J})\nonumber\\
                   &=& \frac{1}{2\sqrt{3}}e_{a}^{\phantom{a}I}e_{b}^{\phantom{a}J}
                       \epsilon_{IJKLM}F^{KL}\ell^{M} \, ;
\eea
it is not difficult to see that the pull-back of this expression to $\Delta$ vanishes
because of the IH condition (\ref{pullback1}) on $\bm{F}$ and the pullback expression
(\ref{pullbackofcoframe}) for $e$.  Thus we again find that
\bea
\nabla_{\!\underleftarrow{a}}\ell_{b}\approx0 \, ,
\eea
and with (\ref{connectionondelta}) it follows that $\omega=0$.  The existence of a
Killing spinor on an IH in five-dimensional EMCS theory implies that the connection
one-form is zero, as was found for the four-dimensional EM theory.

\subsection{Interpretation}

We come to the following conclusion.  The necessary and sufficient condition for
supersymmetry is that there exists a Killing spinor $\psi$ in four dimensions that
satisfies the conditions (\ref{cond1}) and (\ref{cond2}), or a Killing spinor
$\epsilon$ in five dimensions that satisfies (\ref{cond3}).  For an IH these
conditions imply that the induced (normal) connection has to vanish.  This implies
that
\bea
\kappa_{(\ell)} = \omega_{a}\ell^{a} = 0
\quad
\mbox{and}
\quad
\tilde{\omega}_{a} = \omega_{a} + \kappa_{(\ell)}n_{a} = 0 \; .
\eea
The fact that $\omega=0$ means that the IH is nonrotating provided that there exists
a global rotational Killing field $\varphi$ on $\mathcal{M}$ whose restriction to
$\Delta$ is $\phi$.  Therefore we conclude that \emph{an IH is supersymmetric only if
it is extremal and nonrotating}.  These conditions further imply that the topology
constraint (\ref{curvaturecondition}) for SIHs is given by
\bea
\oint_{\mathbb{S}^{D-2}}\bm{\tilde{\epsilon}}\mathcal{R}
= 2\oint_{\mathbb{S}^{D-2}}\bm{\tilde{\epsilon}}(T_{ab}\ell^{a}n^{b} + \Lambda) \; .
\eea
We see that there are two possibilities for the topology of a SIH (when $\Lambda\geq0$).
If $\Lambda>0$ and $T_{ab}\ell^{a}n^{b}\geq0$ or if $\Lambda\geq0$ and $T_{ab}\ell^{a}n^{b}>0$,
then $\oint_{\mathbb{S}^{D-2}}\bm{\tilde{\epsilon}}\mathcal{R}>0$.  In this case the SIH is of
positive Yamabe type.  On the other hand, if $\Lambda=0$ and $T_{ab}\ell^{a}n^{b}=0$ then
$\oint_{\mathbb{S}^{D-2}}\bm{\tilde{\epsilon}}\mathcal{R}=0$.  In this case the SIH can have
torus topology.

In four dimensions, using the IH conditions and the Maxwell field equations and employing the
methodology of Lewandowski and Pawlowski \cite{lewpaw1}, it can also be shown that the quantity
$T_{ab}\ell^{a}n^{b}$ has to be constant over the horizon cross sections and therefore implies
that $\mathcal{R}$ is constant.  Unfortunately, it does not appear that this result holds in
five dimensions because the quantity $T_{ab}\ell^{a}n^{b}$ may vary on the horizon in general.
However, if one could show that $T_{ab}\ell^{a}n^{b}$ cannot vanish on the horizon, then this
would be significant as it would rule out the torus topology.

The quasilocal picture that we have presented is in excellent agreement with the results
that are known for stationary spacetimes \cite{gkltt,gmt}.  In that context a
supersymmetric black hole also contains an extremal and nonrotating horizon.
Extremality is a consequence of the BPS bounds being saturated, which then implies that
there exists a Killing spinor.  Nonrotation is then a consequence of the fact that a
vector cannot be constructed in the neighbourhood of a supersymmetric Killing horizon
that is spacelike, as can be seen from the algebraic conditions (\ref{tnvector1}) and
(\ref{tnvector2}).  Therefore the spacetime of such a black hole cannot contain an
ergoregion, which means that the horizon must be nonrotating.

In five dimensions, there are two supersymmetric solutions with the property that the
bulk electromagnetic field contains angular momentum while the horizon is nonrotating.
These are the BMPV black hole \cite{bmpv} and the EEMR black ring \cite{eemr}.  The
BMPV black hole has one independent rotation parameter.  This corresponds to a SIH
with one angular momentum given by
\bea
\mathcal{J} = \frac{1}{8\pi G_{5}}\oint_{S^{3}}
              (\phi \lrcorner \bm{A})\bm{\Phi} \; .
\eea
The spacetime of the BMPV black hole is described by a nonrotating spherical horizon
with angular momentum stored in the electromagnetic fields.  The angular momentum of
the horizon is nonzero \cite{gmt}.  Therefore some fraction of $\mathcal{J}_{\rm EM}$
is on $\Delta$.  In addition, the distribution of angular momentum of the spacetime
is such that there is a negative fraction behind the horizon as well.  The net result
is that the horizon geometry is that of a squashed three-sphere.  These interesing
properties do \emph{not} contradict the idea that an IH is the inner boundary of the
manifold.  On the contrary, IHs with arbitrary distortions and rotations in their
neighbourhoods have been recently studied using multipole moments \cite{aepv}.  When
the angular momentum of the BMPV black hole is zero the solution reduces to the
extremal RN solution in isotropic coordinates.

The EEMR black ring solution is the generalization of the BMPV black hole solution
to the case where two independent rotation parameters are present.  This corresponds
to a SIH with two angular momenta given by
\bea
\mathcal{J}_{j} = \frac{1}{8\pi G_{5}}\oint_{S^{1}\times S^{2}}
                  (\phi_{j} \lrcorner \bm{A})\bm{\Phi}
\quad
(j\in\{1,2\}) \; .
\eea
In addition, a black ring can also have dipole charges which are naturally defined
on the horizon \cite{cophor,astrad,emprea2}.  For an IH with ring topology a definition
for the dipole charge $\mathcal{P}$ can be realized by integrating the electromagnetic
field strength plus the CS contribution over $S^{2}$:
\bea
\mathcal{P} = \frac{1}{2\pi}\int_{S^{2}} \star \bm{\Phi} \; .
\eea
Charges of this type appear in the first law for a black ring \cite{cophor,astrad}.
However, this is not the case with the first law (\ref{firstlaw}) that we derived.
This is probably due to the fact that the dipole charges $\mathcal{P}$ are associated
with the presence of \emph{magnetic charge}, which we have not incorporated into our
current framework.  If we include magnetic charge, then we should find an additional
term in the first law of the form $\Upsilon\delta\mathcal{P}$ with $\Upsilon$ the
dipole potential at the horizon.

In this paper we focused on null Killing spinors that are defined at the horizon itself.
However, the results obtained here will not be affected for black holes if we assume
that the spinors are defined globally.  We note that there are many other solutions with
such spinors that are defined for the entire spacetime, which do not describe black holes.
These are known as pp-waves (plane-fronted gravitational waves with parallel rays) --
spacetimes with vanishing expansion, shear and twist, and are a subset of a wider class
of solutions that share this property, known as Kundt spacetimes.  For details, see e.g.
\cite{skmhh}.  Given that a spinor is defined globally means that $\ell$, which is
hypersurface orthogonal everywhere, is also defined globally.  Moreover, these spacetimes
can always be foliated by non-expanding horizons \cite{plj}.  In the context of IHs,
we can therefore regard supersymmetry as a condition that fixes the scaling of $\ell$
because $\omega=0$ (independently of the IH condition $[\pounds_{\ell},\mathcal{D}]=0$)
and selects a preferred foliation of the manifold.

\section{Prospects}

In this paper we presented an extension of the IH framework to four-dimensional $N=2$
supergravity and five-dimensional $N=1$ supergravity.  In particular, we derived the
local version of the zeroth and first laws of black-hole mechanics for general WIHs
on the phase space of the $D$-dimensional EMCS theory, and showed that SIHs have to
be extremal and nonrotating.  In the present work we focused mainly on mechanics of
WIHs and the geometrical and topological constraints onto IHs that are imposed by
supersymmetry and the field equations.

There are a number of classical applications of IHs to supergravity black holes that
can be explored.  Here we briefly discuss three problems that are worth investigating.
\begin{itemize}

\item
\emph{Asymptotically ADS spacetime.}
Currently there is a lot of interest in the ADS/conformal field theory (ADS/CFT)
correspondence \cite{maldacena,witten2}, and in particular in finding (supersymmetric)
black hole solutions in ADS spacetime \cite{gutrea,klr1,klr2,kunluc}.  It would
therefore be worth while to study IHs of five-dimensional EMCS theory in asymptotically
ADS spacetime. This would first require extending the covariant phase space that was
constructed here to the phase space of Ashtekar \emph{et al} \cite{apv}, and imposing
the appropriate fall-off conditions onto the fields at the boundary at infinity.  In
addition, the Killing spinor identity would need to be modified due to the presence of
the cosmological constant, thus leading to a modification of the covariant derivative
of the null normal.  This would give the supersymmetry conditions for IHs in
asymptotically ADS spacetime.  It should be noted, however, that topological issues
would need to be approached very carefully in this extension.  This is because the
topology of the black hole event horizon can be affected by the topology of the
boundary at infinity \cite{gsww}.  This suggests that an investigation of the
implications of topological censorship \cite{fsw}, if any, on IHs in asymptotically
ADS spacetime is necessary.  Nevertheless, the resulting framework could provide us
with new insights into the gravitational aspects of the ADS/CFT correspondence.  

\item
\emph{BPS bounds.}
The general method for deriving the BPS bound for stationary spacetimes is to construct
an expression for the energy of the spacetime using spinor identities and the Einstein
field equations.  This method was pioneered by Witten \cite{witten1} and Nester
\cite{nester} to prove the positive energy theorem.  The method was applied in four
dimensions \cite{gibhul} and in five dimensions \cite{gkltt} to calculate the BPS bounds
for the corresponding spacetimes.  How can one derive these bounds for IHs?  The bounds
are saturated when the spinors are supercovariantly constant, which is associated with
extremality.  This suggests that the extremality condition (\ref{characterization}) can
be used for IHs.  This is straight-forward to do for undistorted horizons.  Let us
consider the four-dimensional EM theory for illustration.  Here the contraction
$T_{ab}\ell^{a}n^{b}$ is the square of the electric flux $E_{\perp}$ crossing the
surface \cite{abf2}.  For any IH this quantity is constant over $S^{2}$ and can
therefore be moved outside the integral.  The result can be used to relate the charge
$Q$ on the horizon to its surface area $A$ via $Q=E_{\perp}A/(4\pi)$.  For the RN
solution one finds that $\eta=Q^{2}/R^{2}-1\leq0$ with $R=\sqrt{A/(4\pi)}$ the areal
radius \cite{boofai2}.  When the surface gravity vanishes $\eta=0$ and $Q=R=M$ with
$M$ the mass.  This is the condition for supersymmetry in four dimensions.  The
situation is not as obvious for distorted horizons in five dimensions.  This is because
the contraction $T_{ab}\ell^{a}n^{b}$, which for EMCS theory is again the square of the
electric flux, may not be constant over the horizon cross sections in general.  However,
for the BMPV black hole in particular we know that the cross sections are $S^{3}$ which
have constant curvature, and therefore $E_{\perp}^{2}$ is constant on $\Delta$.  From
here, a charge-areal radius relation follows along the same lines as the derivation
that was outlined above for the RN black hole in four dimensions.

\item
\emph{Supersymmetry and black-hole uniqueness.}
In four dimensions, it was shown \cite{lewpaw1} that the IH constraints for extremal
IHs of EM theory are satisfied iff the intrinsic geometry of the horizon coincides
with that of the extremal Kerr-Newman (KN) solution.  An extension of that analysis
to IHs of five-dimensional EMCS theory would be of interest, particularly because it
would provide a method for deriving the geometries of the corresponding extremal IHs.
This would complement a recently developed method \cite{klr3,astyav} for deriving the
\emph{near-horizon} geometries of extremal black holes.  While speculating on the
local uniqueness theorems in five dimensions we need to keep in mind that black holes
in five dimensions are much less constrained than in four dimensions, mainly because
in five dimension there are two possible topologies ($S^{3}$ and $S^{1}\times S^{2}$),
and also because there are two independent rotation parameters.  As a consequence of
this richer structure, it is possible that two distinct black holes in five dimensions
can have the same asymptotic charges \cite{emprea1}.  This is a striking example of
black-hole nonuniqueness in higher dimensions.  Nevertheless, uniqueness has been
established for supersymmetric black holes in five dimensions \cite{reall}.  Therefore
it is expected that the five-dimensional analogues of the local uniqueness theorem of
\cite{lewpaw1} do exist, but for SIHs.  We also note that, while supersymmetry
constrains the geometry (i.e. connection one-form), the dominant energy condition and
the Einstein field equations are still required to constrain the topology.  Therefore
we expect that there should be a unique horizon geometry for a given topology.  For
example, if the topology of a SIH is $S^{1}\times S^{2}$, then the geometry should
coincide uniquely with the induced metric and vector potential of the EEMR black ring
solution.  We also expect that, if the topology is $S^{3}$, then the geometry should
coincide uniquely with that of the BMPV black hole \emph{in general}, and the extremal
RN black hole as a limiting case when the angular momentum of the Maxwell fields vanishes.
It would be of considerable interest to try solving the IH constraints for a SIH when
$T_{ab}\ell^{a}n^{b}=0$ (at the horizon); the resulting geometry would provide the
first explicit solution of a supersymmetric black hole with toroidal topology.

\end{itemize}
The above are just three examples of how the IH framework can be applied to investigate
the classical gravitational aspects of supergravity and superstring theory.  The extension
of IHs to supergravity originated with the objective of employing Hamiltonian methods to
study aspects of the ADS/CFT correspondence.  The extension of the work presented here to
asymptotically ADS spacetimes and derivation of the corresponding Hamiltonian is currently
in progress \cite{liko2}.  This will provide the point of departure for the study of quantum
fields on null surfaces, with applications to black hole statistical mechanics in the ADS/CFT
correspondence.

\section*{Acknowledgements}

This work was supported by the Natural Sciences and Engineering Research Council
of Canada through a PGS D3 Scholarship (TL) and a Discovery Grant (IB).


\end{document}